\documentclass[10pt]{iopart}

\usepackage{iopams}  
\usepackage{graphicx}
\usepackage{cite}
\begin{document}

\title[]{Runaway electron deconfinement in SPARC and DIII-D by a passive 3D coil}

\author{V.A. Izzo$^1$, I. Pusztai$^2$, K. Särkimäki$^3$, A. Sundström$^2$, D. Garnier$^4$, D. Weisberg$^5$, R.A. Tinguely$^4$,  C. Paz-Soldan$^6$, R.S. Granetz$^4$, R. Sweeney$^4$}

\address{$^1$Fiat Lux, San Diego, CA 92101,USA}
\address{$^2$Department of Physics, Chalmers University of Technology, SE-41296 G\"{o}teborg, Sweden}
\address{$^3$Max Planck Institute for Plasmaphysics, 85748 Garching, Germany}
\address{$^4$Plasma Science and Fusion Center, Massachusetts Institute of Technology, Cambridge, MA 01239, USA}
\address{$^5$General Atomics, San Diego, CA 92121, USA}
\address{$^6$ Department of Applied Physics and Applied Mathematics, Columbia University, New York, NY 10027, USA}

\ead{val@fiatlux.energy}
\vspace{10pt}

\begin{abstract}
The operation of a 3D coil--passively driven by the current quench loop voltage--for the deconfinement of runaway electrons is modeled for disruption scenarios in the SPARC and DIII-D tokamaks. Nonlinear MHD modeling is carried out with the NIMROD code including time-dependent magnetic field boundary conditions to simulate the effect of the coil. Further modeling in some cases uses the ASCOT5 code to calculate advection and diffusion coefficients for runaway electrons based on the NIMROD-calculated fields, and the DREAM code to compute the runaway evolution in the presence of these transport coefficients. Compared with similar modeling in Tinguely, et al [2021 Nucl. Fusion 61 124003], considerably more conservative assumptions are made with the ASCOT5 results, zeroing low levels of transport, particularly in regions in which closed flux surfaces have reformed. Of three coil geometries considered in SPARC, only the $n=1$ coil is found to have sufficient resonant components to suppress the runaway current growth. Without the new conservative transport assumptions, full suppression of the RE current is maintained when the TQ MHD is included in the simulation or when the RE current is limited to 250kA, but when transport in closed flux regions is fully suppressed, these scenarios allow RE beams on the order of 1-2MA to appear. Additional modeling is performed to consider the effects of the close ideal wall. In DIII-D, the current quench is modeled for both limited and diverted equilibrium shapes. In the limited shape, the onset of stochasticity is found to be insensitive to the coil current amplitude and governed largely by the evolution of the safety-factor profile. In both devices, prediction of the q-profile evolution is seen to be critical to predicting the later time effects of the coil.   
\end{abstract}

%
%
%
%
\ioptwocol

    \section{Introduction}\label{sec:intro} 
The high current tokamaks of the future are especially susceptible to the risk of multi-MA beams of runaway electrons carrying 10s of $\rm MeV$ energies, which could damage in-vessel components or quench superconducting coils. This can occur following a thermal quench (TQ), when a sudden loss of confinement cools the plasma and raises the electric field ($E$) above the critical electric field ($E_{\rm crit}$) for runaway electron (RE) production \cite{Dreicer1959}. In the $E>E_{\rm crit}$ regime, several mechanisms of runaway production are possible, but the secondary or \textit{knock-on avalanche} mechanism is proportional to the exponential of the initial plasma current \cite{rosenbluth:1997}, so that almost any primary RE source can provide enough seed for complete conversion of the thermal current when the tokamak flat-top $I_p$ is $\sim 10\,\rm MA$. 

A variety of strategies have been investigated to avoid, suppress, or mitigate runaway electrons. Experiments aimed at mitigating tokamak disruptions with massive material injection (gas or pellets) have long sought unsuccessfully to exceed the so-called Rosenbluth density to maintain $E<E_{\rm crit}$ even at post-thermal quench temperatures \cite{Whyte2002,Granetz2007,Commaux2010}. Injection of either high-$Z$ (Ne, Ar, Kr, Xe) \cite{reux2015runaway,shiraki2018dissipation} or low-$Z$ (He, ${\rm D}_2$) \cite{shiraki2018dissipation,Hollmann2013} material into an existing runaway electron beam has also been pursued, with ${\rm D}_2$ injection showing strong promise experimentally for benign termination of a runaway electron beam by a kink instability, supported by extended-MHD modelling \cite{reux2021demonstration,Paz-Soldan2021,Liu2019}, which has been extended to ITER scenarios \cite{Liu_2022}. Strategies to prevent the formation of a mature RE beam involve enhancing transport of the seed REs in physical or momentum space, e.g., by stochastic magnetic fields \cite{Lehnen2008,Gobbin2018,Commaux2011,Svensson,Mlynar2019,Liu2020} or wave-particle interactions \cite{Guo2018,PhysRevLett.120.265001}, to produce a loss rate that exceeds the avalanche growth rate. 

MHD fluctuations produced by instability during the TQ can play a role in deconfining seed REs, but re-healing of flux surfaces early in the current quench (CQ) allows any remaining seeds plenty of opportunity to exponentially amplify in high current devices. Modeling has also predicted a very unfavorable $R^3$ size-scaling for RE confinement during the TQ \cite{Izzo2011}, so that large devices like ITER may not benefit much from these naturally occurring losses. Several tokamaks have explored the possibility of enhancing RE seed losses in the TQ and early CQ phase of the disruption with the application of external 3D fields produced by actively driven coils. This technique showed some positive effect in DIII-D diverted plasmas \cite{Commaux2011}, as well as in TEXTOR \cite{Lehnen2008}. ASDEX experiments, interpreted with modeling using MARS-F and ORBIT, found that the most effective RMPs were those that enhanced drift of high energy REs in the edge region \cite{Gobbin_2021}. Experiments in J-TEXT and MHD modeling  \cite{Chen2016} found that, in some cases, these 3D fields could enhance RE confinement. Observations in DIII-D also provide evidence of the role of high-frequency kinetic instabilities in enhancing RE losses \cite{Lvovskiy2019}. 

A strategy to enhance losses of REs in the CQ with stochastic fields produced by a passively-driven 3D conductor was first proposed by Boozer \cite{boozer2011two,smith2013passive}. The concept relies on the fact that the large loop voltage during the CQ can drive considerably more current in a non-asymmetric conductor than would be practical with a coil actively driven by power supplies. An added virtue is that initiation of the 3D-coil current does not rely on a disruption-prediction algorithm in a control system but on the more dependable laws of physics. Modeling of a passive runaway electron mitigation coil (REMC) for the SPARC tokamak \cite{creely_2020,sweeney_2020,Rodriguez-Fernandez2022} was first reported by Tinguley, \textit{et al.} \cite{Tinguely_2021}, and combined calculations from a series of four codes---COMSOL \cite{comsol} for the coil fields, NIMROD \cite{sovinec:2004} for the nonlinear MHD plasma response, ASCOT5 \cite{hirvijoki2014ascot} for RE transport coefficients, and DREAM \cite{Hoppe_DREAM-2021,Svenningsson2021} for RE generation and total current evolution---to predict total suppression of RE beam formation in SPARC with a (toroidal mode number) $n=1$ passive coil. Those results relied primarily on a single NIMROD simulation which included only the CQ MHD fluctuations driven by the REMC. Here we present additional NIMROD modeling of SPARC disruption scenarios, which include the effects of TQ MHD and also examine how the CQ temperature and the location of the perfectly conducting wall (each of which will alter the CQ duration), and the maximum coil current impact the results. The confinement of a small number ($\sim 10,\!000$) of RE drift-orbits calculated directly within NIMROD is used as a simple assessment of RE confinement under various scenarios, but this method does not collect the level of statistics vs.~time, space, energy and pitch used to obtain transport coefficients with ASCOT5. In two additional cases, the combined modeling including ASCOT5 and DREAM is repeated. The $n=1$ coil is also compared with $n=2$ and $n=3$ coil geometries to establish the basis for the selection of the $n=1$ configuration.

Optimization and modeling of a passive 3D coil for DIII-D \cite{Fenstermacher2021} was reported by Weisberg, \textit{et al} \cite{Weisberg2021}. In that study, the 3D fields from the optimized coil were handed off to the MARS-F code to calculate the linear plasma response and RE drift-orbits for a time midway through the DIII-D CQ. The modeling showed RE loss fractions ranging from ~30-70\% (for an initial distribution that is uniform in the poloidal flux radial coordinate, $\psi$), depending on the coil configuration, coil current, and $q$-profile of the equilibrium. Note that this loss fraction can not be directly translated to an equivalent reduction in final RE current. Here we use the same 3D coil fields to perform time-dependent, nonlinear simulations of DIII-D CQ scenarios with the passive coil, also including RE drift orbit-calculations. A limited equilibrium similar to the MARS-F study is considered, as well as a diverted equilibrium to connect to the SPARC modeling. In both cases we study only the MHD produced by the coil during the CQ, although the effects of TQ MHD on RE confinement have been modeled with NIMROD previously for DIII-D in both diverted \cite{Izzo2011,Izzo2012,James2011} and limited \cite{Izzo2011,James2011} plasma shapes.      

    \begin{figure}[ht]
    \includegraphics[width=0.48\textwidth]{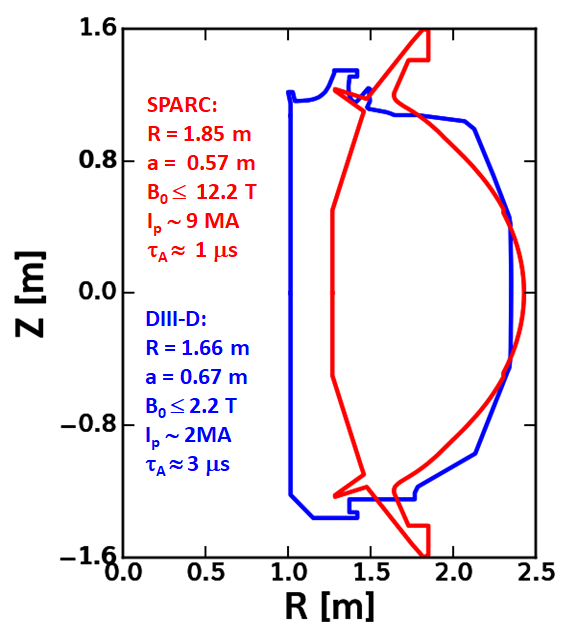}
    \caption{The design of the SPARC device is similar in size to DIII-D (first-wall shapes are shown), but with much higher field and current density. Approximate Alfv\'en times use the length of the magnetic axis and and assume a pure deuterium plasma of $4\times10^{20}/m^3$ in SPARC and $10^{20}/m^3$ in DIII-D. 
    }
    \label{fig:Shapes}
    \end{figure}

The SPARC \cite{creely_2020,Rodriguez-Fernandez2022} and DIII-D \cite{Fenstermacher2021} tokamaks are quite comparable in size and aspect ratio (shown on the same scale in Figure \ref{fig:Shapes}), but differ considerably in magnetic field strength and total plasma current. Important effects of the higher field and current in SPARC for the present study include the much ($\sim$ 40 million times) larger expected RE avalanche multiplication during the CQ, as well as the shorter Alfv\'en time, which will lead to faster growth of both ideal and resistive MHD modes. The difference in avalanche multiplication tends to make RE plateau formation in SPARC insensitive to the initial seed, while DIII-D is sensitive to the seed, and to the loss of some fraction in the TQ or early CQ phase \cite{Lvovskiy2018,Izzo2012,James2011}. Although at smaller size and somewhat lower maximum current, SPARC resembles ITER insofar as both occupy the seed-insensitive regime of avalanche multiplication. At present, SPARC and DIII-D plasmas also differ existentially, where DIII-D has been operating since the mid-1980's \cite{PhysRevLett.59.1432} and SPARC operation is planned to begin in 2025 \cite{chrobak2021overview}, however the planned REMC coils for both devices presently exist only on paper. Some important parameters for the two devices are compared in Figure \ref{fig:Shapes}.

    \section{SPARC REMC modeling}\label{sec:sparc}

The single SPARC REMC simulation described in detail in Ref.~\cite{Tinguely_2021} involved an artificially rapid thermal quench produced by temporarily increasing the perpendicular thermal conduction to bypass MHD activity during the TQ and compute only the MHD plasma response to the $n=1$ REMC. Among the additional cases modeled here are: a simulation with a more realistic TQ produced by impurity radiation, leading to TQ-induced MHD activity; a case at higher TQ temperature (slower CQ); and a case with the maximum coil current clamped at $250\,\rm kA$. In order to study the effect of the close conducting wall, we also compare TQ simulations with no REMC for two different wall locations. First, we begin with a direct comparison of three coil configurations, namely $n=1$, $n=2$, and $n=3$, highlighting the physics basis for down-selecting to the $n=1$ configuration for the SPARC REMC design and ongoing modeling efforts.

\subsection{Comparison of coil geometries}
 
Three separate NIMROD simulations of CQ MHD induced by $n=1$, $n=2$, and $n=3$ REMC coils in SPARC were carried out to compare the performance of these three coil geometries, where the three coil designs consist, respectively, of two, four and six vertical legs along the outboard wall, connected by alternating upper and lower horizontal legs, as illustrated in Figure 12 of Ref.~\cite{sweeney_2020}. The coils are labeled by their dominant toroidal component, but the magnetic fields produced by each contain more than one Fourier harmonic, and all components of the coil fields up to the resolution of the modeling (n=0-10) are included. The second largest mode for each coil is the $3n$ component, which is explained simply by the fact that the Fourier decomposition of a square wave contains only the odd-integer multiples of the dominant mode. A secondary effect is the fact that the SPARC wall is designed with nine toroidal segments, so that only the $n=3$ coil has horizontal segments all of equal length, producing a slightly altered spectrum of sub-harmonics for the $n=1$ and $n=2$ coils. Note that the $n=3$ coil geometry was explored first and was modeled with an earlier SPARC equilibrium than the other two coils, which will be seen to result in a shorter $L/R$ time (where $L$ and $R$ are the plasma inductance and resistance) and faster avalanche growth rate. This is not expected to significantly influence the overall effectiveness of the coils. 

In each of these simulations, the CQ begins immediately following a fast, artificial TQ, produced by briefly increasing the perpendicular thermal conduction to $\kappa_{\perp}=4.0\times 10^4 \,\rm m^2/s$  until nearly all the plasma stored energy is lost in a manner that triggers no associated MHD instabilities. Once the current begins to decay, the REMC fields grow and nonlinear mode growth in the CQ phase is triggered in response to these 3D perturbations. The coil perturbations are applied by directly imposing the normal component of the coil vacuum magnetic fields at the simulation boundary (approximately the limiter location), with the coil current amplitude varied in proportion to the time changing plasma current, according to:
\begin{equation}
I_{\rm REMC}=I_{\rm max}\left(1 - \frac{I_p}{I_{p,t=0}} \right),
\end{equation}
where the maximum coil current $I_{max}$ is prescribed in NIMROD based on the results of vacuum field calculations from COMSOL as described in \cite{Tinguely_2021}. The simple linear relationship of the REMC current and plasma current used in NIMROD is indicative of the low resistivity of the coil and justified based on the output of the COMSOL calculations. In addition to the time-varying normal component of $\mathbf{B}$, these ideal-wall calculations also have tangential electric fields imposed explicitly at the boundary for consistency with Faraday's law. The freedom in the choice of solution for the electric field (the gradient of scalar voltage) physically corresponds to the locations where magnetic flux connecting the inward and outward normal fields can soak into the volume, which would in reality be determined by physical properties of the wall, such as the locations of insulating gaps. For numerical convenience, purely poloidal electric fields with purely toroidal derivatives in the curl are chosen, allowing simple analytical evaluation given NIMROD's Fourier representation of the fields in the toroidal direction.

\begin{figure}[ht]
    \includegraphics[width=0.48\textwidth]{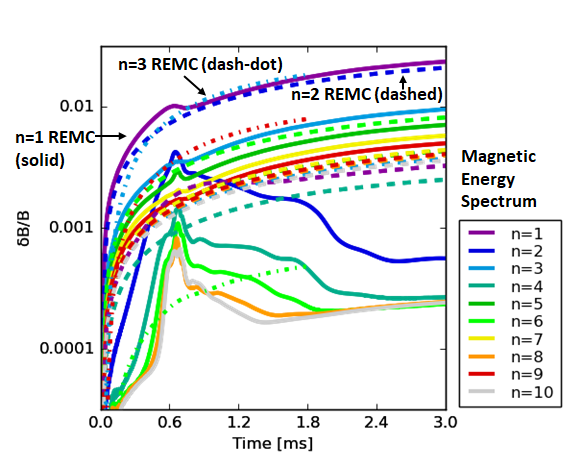}
    \caption{For three SPARC coil geometries the amplitudes of the dominant modes are comparable, but nonlinear mode growth is most apparent for the $n=1$ coil. The magnetic energy spectrum (toroidal modes $1$-$10$, differing colors) is plotted in units of $\delta B/B$ (square root of mode energy over n=0 energy), for $n=1$ coil (solid), $n=2$ coil (dashed), and $n=3$ coil (dash-dot).     
    }
    \label{fig:SPARC_modeSpectra}
    \end{figure}

The nonlinear time evolution of the magnetic energy spectrum for all three coil configurations is plotted in Figure \ref{fig:SPARC_modeSpectra}. The dominant driven mode for each coil configuration reaches comparable relative amplitude, with $\delta B/B \gtrsim 10^{-2}$ (using the square root of the ratio of mode magnetic energy to n=0 magnetic energy as a global measure). For the $n=3$ REMC, the next largest component is $n=9$, followed by $n=6$, with no other mode amplitudes exceeding noise level. The $n=2$ REMC has a subdominant $n=6$ harmonic, followed by several smaller harmonics at comparable amplitude. A slight non-monotonic behaviour of the $n=1$ mode can be observed between $0.5$ and $1\,\rm ms$ for the $n=2$ REMC, indicating some nonlinear growth and saturation of that mode, but not at large amplitude. The $n=1$ REMC only perturbs the odd toroidal mode numbers directly, but differs significantly from the other configurations in that it can been seen to drive a large nonlinear response, resulting in growth and saturation of all modes at around 0.65 ms, with the un-driven $n=2$ mode transiently becoming the second largest. This $n=1$ REMC simulation was analysed in more detail in Ref. \cite{Tinguely_2021} and is included here for direct comparison with the other coil configurations.

In subsequent sections, the ASCOT5 code will be used to accurately obtain transport coefficients from NIMROD calculated fields for some simulations. However, to estimate RE losses within NIMROD, drift orbits of 56,240 passing electrons are initiated uniformly as a function of poloidal flux at random poloidal and toroidal locations, with energies between $0$ and $50 \,\rm MeV$, and pitch (defined as $p_{\perp}/p$) of $0-0.5$, at the start of each simulation. In this calculation, the energy of each test-particle is held fixed. Although the NIMROD test-particle module has the capability to evolve energy according to the electric field, small-angle collisions, and radiation terms \cite{Izzo2021}, without a pitch-angle scattering term, the correct evolution in momentum space would not be captured. Instead, the initial distribution is uniform over the energy and pitch range specified in order to allow a simple comparison of how different regions of momentum space are confined. As orbits are lost to the simulation boundary a global loss rate is calculated and compared with an approximate form of the avalanche growth rate (assuming $E \gg E_{\rm crit}$) \cite{rosenbluth:1997}, using the time derivative of the poloidal flux for a global estimate of the electric field.
\begin{equation}
\gamma_{RA}\approx\frac{eE}{2mc\ln\Lambda}\approx\frac{e\dot{\psi}_{pol}}{4\pi Rmc\ln\Lambda}.
\end{equation}
We reiterate that in the subsequent calculations of total RE evolution with DREAM, the simple NIMROD test-particle loss estimates are not used, only the NIMROD magnetic field evolution, which is used to obtain transport coefficients more accurately with ASCOT5.

\begin{figure}[ht]
    \includegraphics[width=0.48\textwidth]{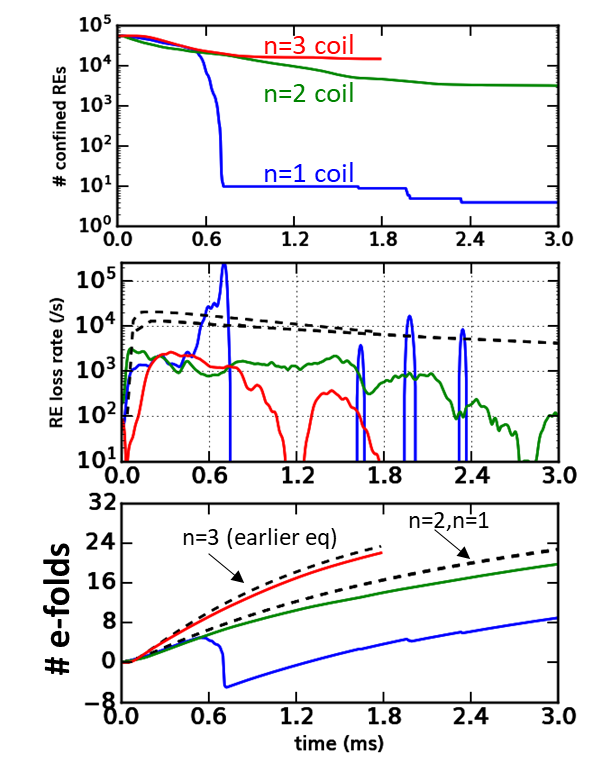}
    \caption{Only the $n=1$ coil produces an RE loss rate that exceeds the avalanche growth rate. (a) Total number of confined REs vs. time. (b) RE loss rates ($\dot{N}_{RE} / N_{RE}$, solid) compared with avalanche growth rates (dashed)--the slightly higher growth rate is for the $n=3$ simulation with an earlier equilibrium; the $n=1$ and $n=2$ growth rates overlay. (c) Estimated number of avalanche e-folds from integrated growth rate minus loss rate (solid) compared with integrated growth rates only (dashed). In the n=1 case, RE losses almost cease with a small RE population confined in a central island, allowing the avalanche growth term to again dominate at late times.
    }
    \label{fig:SPARC_n123}
    \end{figure}

With RE orbits initiated uniformly over the cross-section, all three coil configurations result in the loss of a majority of the test-particles (Figure \ref{fig:SPARC_n123}). The final distributions show losses of nearly all electrons outside of a confined central region, irrespective of energy. With the largest retained population, the $n=3$ coil simulation shows a slight trend of better confinement for higher energy electrons (i.e., $16\%$ more retained REs above 40MeV than below 10MeV). The smaller retained populations for the $n=2$ and $n=1$ coils do not show a clear trend in energy, although statistics for the retained population with the $n=1$ coil are particularly poor. However, the $n=2$ and $n=3$ coil simulations not only retain an appreciable fraction of the initial test-particles ($5\%$ and $30\%$, respectively), due to incomplete flux surface destruction in the core, but neither case ever exhibits a global RE loss rate exceeding the estimated avalanche growth rate. Consequently, only a small change is estimated to the large number of avalanche e-folds expected in SPARC, where such a small reduction is likely to have no marked effect on the final RE current given any appreciable seed current. The $n=1$ coil, by contrast, deconfines nearly all the test-particles ($0.02\%$ remain confined in this case, all orbiting very close to the magnetic axis, where a small flux tube persists), and also causes RE losses that are transiently faster than the avalanche growth rate by more than an order of magnitude. It should be emphasized that the loss rates obtained from NIMROD, with a single test population launched once at $t=0$ (in some cases relaunched later in time), provide only an estimate for simple comparison between cases. For instance, the sudden drop-off in the loss rate in the $n=1$ coil simulation is primarily because the REs have already been lost from all but a very small region, not because good flux surfaces have reappeared in the unconfined regions. The analysis published in Ref. \cite{Tinguely_2021}, which used the ASCOT5 code to launch RE orbits at many time slices (with a range of energies and pitches) and obtain much more accurate transport coefficients, predicted total suppression of the RE current in this $n=1$ simulation. In effect, the NIMROD drift-orbit calculations can be used to rule out the $n=2$ and $n=3$ coils, while the more detailed multi-code analysis more rigorously qualifies the $n=1$ configuration. Additional NIMROD simulations to explore the behavior of the $n=1$ coil under different scenarios and the effects of some approximations in the modeling are presented in the next two sections.  

\subsection{Inclusion of TQ MHD and effects of the ideal wall}

In order to isolate the effects of the coil on the RE confinement, the TQ-triggered MHD activity was ignored in the preceding scenarios. In reality, the MHD modes appearing during the TQ are by themselves expected to produce a significant loss of RE confinement \cite{sweeney_2020}, but the concern remains that without the REMC the flux surfaces may reheal early in the CQ. We now model the scenario in which the TQ is initiated in a more realistic manner by the addition of a large quantity of Ne (as might occur when the massive gas injection system is triggered for disruption mitigation) which radiates the plasma thermal energy in roughly 1 ms and drives unstable MHD modes. We compare the scenario with both the realistic TQ and the $n=1$ coil to the scenario with only the $n=1$ coil, and also to the scenario with a realistic TQ and no coil. 

\begin{figure}[ht]
    \includegraphics[width=0.48\textwidth]{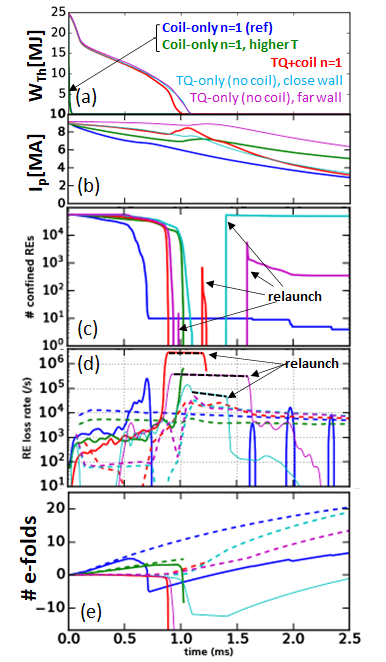} 
    \caption{Evolution of (a) thermal energy and (b) plasma current for simulations with and without the $n=1$ REMC are compared to the $n=1$ reference case (blue) in Figure \ref{fig:SPARC_n123}, including a case with a higher CQ temperature (green), a case including a more realistic TQ scenario (red), and two cases in which only the TQ MHD (cyan, magenta) is modeled for comparison with coil simulations and to understand the effects of the close, ideal wall (cyan has the same wall location as all other cases). Results for RE drift orbits for each case are also plotted: (c) Total number of confined REs vs. time. (d) RE loss rates (solid) compared with avalanche growth rates (dashed). In every case but the reference case, the number of confined REs goes to zero. In some cases a population of REs is repeatedly re-launched until a significant population remains confined. The black dashed lines indicate times between loss of all REs and final relaunch where a loss rate is not calculated. (e) Estimated number of avalanche e-folds from integral of growth rate minus loss rate (solid) compared with integrated growth rates (dashed). 
    }
    \label{fig:SPARC_n1all}
    \end{figure}

The thermal energy and plasma current traces for these simulations, and others to be discussed in this and the next section, are plotted in Figure \ref{fig:SPARC_n1all}. Because the coil-only simulations have a nearly instantaneous TQ, the start of the CQ between these simulations and those with a realistic TQ differs by 1 ms. For the simulation that includes both a realistic TQ and the $n=1$ REMC (red lines), we note that the end of the TQ is slightly shortened compared to cases with no REMC (TQ-only), and that a more prominent $I_p$-spike is observed (as well as a larger peak $n=1$ mode amplitude, not shown), compared both to TQ-only cases (cyan, magenta) and the coil-only cases (blue, green). Following the large $I_p$-spike, a faster current decay is observed, which can be attributed to the drop in inductance associated with the $I_p$-spike. As a consequence, the current decay in the TQ + coil simulation catches up with the coil only case, in spite of the later CQ start time. 

RE losses for the same set of cases are also plotted in Figure \ref{fig:SPARC_n1all}. First, we note that in the cases with the realistic TQ, $100\%$ of the RE test-particles are lost. That is, the TQ by itself, at least transiently, is more effective at deconfining the RE population than the coil by itself. But while the TQ-only simulation reaches a loss rate just above the avalanche growth rate, and comparable to the n=1 coil only simulation, the TQ + n=1 coil simulation reaches a much higher loss rate than any other simulation, exceeding the avalanche growth rate by two orders of magnitude. Also significant is that the losses occur just as the CQ begins, so that while several avalanche e-folds are predicted prior to the rapid loss in the coil-only case, no significant early avalanching of the seed population is expected here. 

In both the TQ-only and TQ + $n=1$ coil simulations, a new test population with an identical distribution to the initial population is launched repeatedly after the initial population is lost, until some fraction remains confined within the time interval of the data output ($5\,\rm  \mu s$). The relaunching is done manually on a coarse, irregular time scale as changes in the magnetic topology are observed; it does not indicate the exact timing of reappearance of regions of confinement, but allows a comparison of the extent to which confined regions have reappeared after similar time intervals. With TQ MHD only, the large majority ($97\%$) of the initial test population is once again confined after $0.3\,\rm ms$ following the loss of all REs. Within a very similar time frame, we find only a tiny fraction ($1\%$) of the new test population being retained in the TQ + $n=1$ coil simulation. The relaunched population always has an initial rapid loss of some fraction of the test-population (high loss rate), followed by a drop off in the loss rate once the REs in the unconfined region have been lost. No global loss rate is calculated between the initial loss of all REs and the first relaunch that retains some fraction of the population. Interpolating the loss rate of the TQ + $n=1$ coil simulation by a straight line across this time interval of unknown confinement yields an estimated reduction of the RE population by $-2500$ e-folds, well off the scale of the plot in part (e) of Figure \ref{fig:SPARC_n1all}.

\begin{figure}[ht]
    \includegraphics[width=0.48\textwidth]{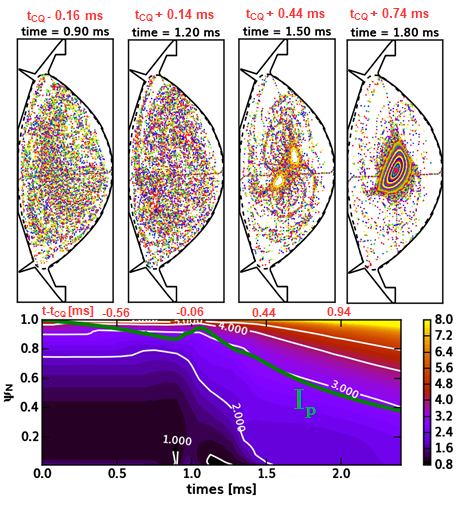}
    \caption{(Top) Field line Poincar\'e plots for the TQ+coil simulation show rapid rehealing of core flux surfaces after 1.5 ms. The solid black line is the SPARC limiter shape and the dashed line is the simulation boundary. (Bottom) Evolution of the $q$-profile (color contours and white lines) shows that after the $I_p$-spike (green line) the $q$ on-axis sharply rises, and the flux surface rehealing begins as $q$ on-axis crosses above two. Times relative to the CQ in red are calculated with respect to the peak of the $I_p$-spike. 
    }
    \label{fig:SPARC_reheal}
    \end{figure}

While the dramatic losses predicted for the TQ + $n=1$ coil simulation bode well for the REMC effectiveness, this simulation is less optimistic than the coil-only simulation in one aspect, which is the timescale of flux-surface re-healing. The large $I_p$-spike that occurs only with the combination of the TQ MHD and the $n=1$ coil is associated with a significant redistribution of the current profile, and in particular, a sharp increase in the on-axis $q$ (which occurs on a slower timescale in the coil-only simulation). As seen in Figure \ref{fig:SPARC_reheal}, as the safety factor on-axis crosses above two, two small islands form and rapidly coalesce into a large region of good flux surfaces. Once $q_0$ exceeds two, a significant region of the core has $2<q<3$ with no $n=1$ resonant surfaces.

In all figures showing Poincar\'e plots, we represent the magnetic field lines only. It is known that high energy REs with large curvature drift can effectively average over some magnetic perturbations and "see" different levels of stochasticity \cite{Mynick1981}, and, although not represented in the plots, this effect is accounted for in the drift orbit calculations in both NIMROD and ASCOT5 (seen for instance in the decreasing transport at large momentum in Ref. \cite{Tinguely_2021}).

For this simulation, we perform the same analysis that was carried out in Ref.~\cite{Tinguely_2021} for the CQ-only simulation.  The runaway electron diffusion and advection coefficients corresponding to the perturbed magnetic field of the NIMROD simulations were calculated by the orbit following code ASCOT5, as described in \cite{Sarkimaki_2016,Sarkimaki_2020}, and are provided as functions of time, radius, and momentum. Note, that in the current work the transport coefficients are post-processed to remove low values of the transport coefficients, especially from regions where the Poincar\'e plot of the magnetic field indicate the presence of intact flux surfaces. These advection and diffusion coefficients are then translated into transport coefficients of the runaway particle density using the method of Ref.~\cite{Svensson}, and used in the fluid mode of the DREAM code to calculate the runaway evolution. The simulation accounts for the primary generation through the Dreicer, hot-tail, tritium decay, and Compton scattering mechanisms, as well as runaway avalanche. The Dreicer rate is calculated using a neural network \cite{Neutral},  the avalanche accounts for partial screening \cite{Hesslow_2019}, and the conductivity used is valid across all collisionality regimes \cite{Sauter99}. 

\begin{figure}[ht]
    \includegraphics[width=1.0\columnwidth]{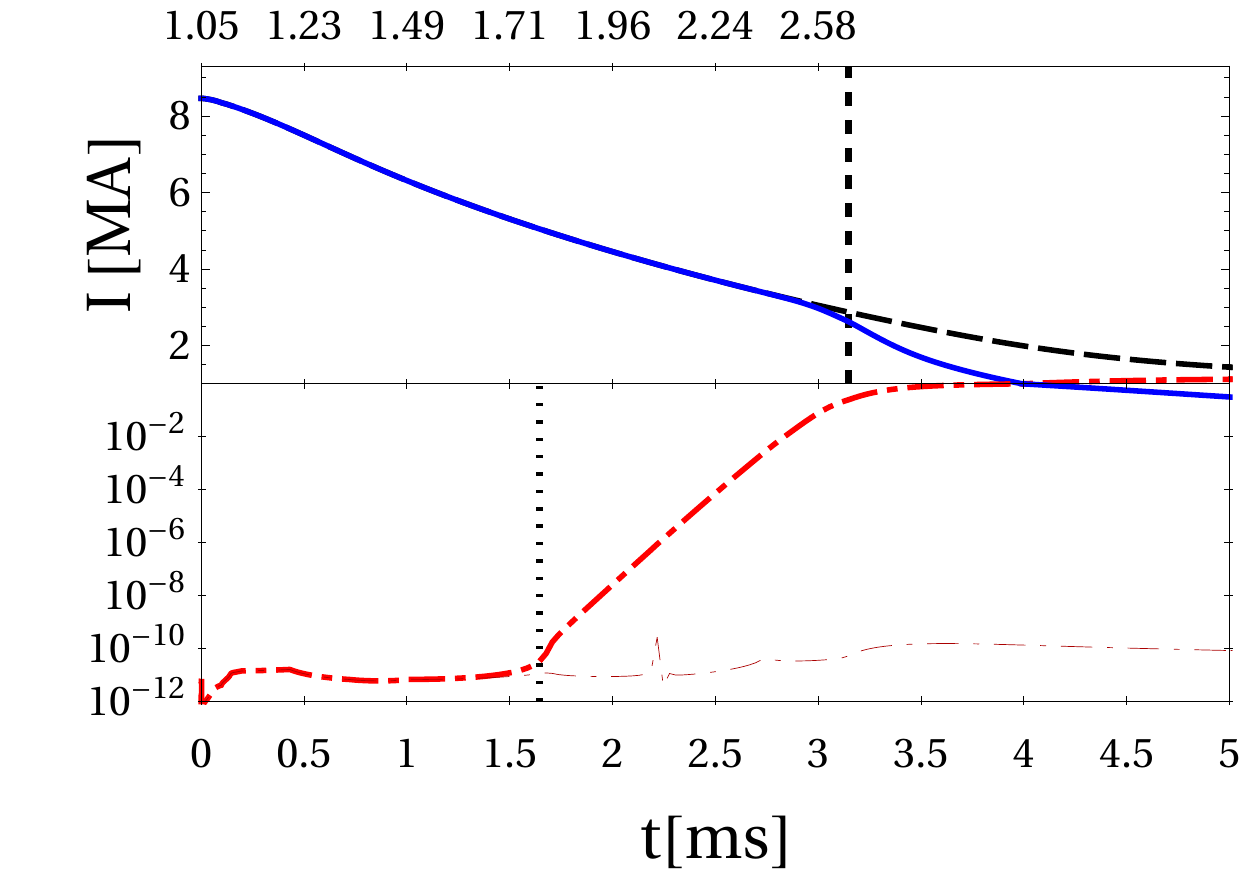}
    \caption{Evolution of the total plasma current (dashed black) and its Ohmic (solid blue) and runaway (thick red dash-dotted) components, calculated with DREAM for the natural TQ+$n=1$ coil case (corresponding to the red curve in Fig.~\ref{fig:SPARC_n1all}). Transport coefficients are fixed after the time indicated by the vertical dashed line in the upper panel. The time of the reformation of good flux surfaces in the core indicated by the vertical dotted line in the lower panel. Mind the linear (logarithmic) scales on the upper (lower) panel. Note that the simulation time in DREAM, shown in the lower horizontal axis, is not the same as that in NIMROD, shown in the upper horizontal axis; the time of the flux surface healing (dotted vertical line) maps to $1.8\,\rm ms$ in NIMROD time (corresponding to the rightmost upper panel of Fig.~\ref{fig:SPARC_reheal}). We also show the runaway current calculated for transport coefficients without post-processing (thin red dash-dotted line).  
    }
    \label{fig:DREAM_TQCQ}
    \end{figure}

In this case the initial current density profile used in DREAM is the partially relaxed profile taken from the peak of the $I_p$-spike in NIMROD (at simulation time $1.05\,\rm ms$). The transport coefficients are mapped to the DREAM simulation as functions of the plasma current, instead of time in the NIMROD simulation, requiring a monotonic $I_p$ variation. The DREAM simulation begins with a $0.096\,\rm ms$ long prescribed exponential temperature decay with a characteristic decay time of $0.05\,\rm ms$, starting from the temperature profile in the NIMROD simulation at $0.9\,\rm ms$. With regards to RE losses, the simulation effectively begins at $1.05\,\rm ms$ of the NIMROD simulation. Note that the RE losses calculated by NIMROD begin prior to $1.05\,\rm ms$; indeed all of the test particles are lost by that time. However, the DREAM simulation shows that following the natural thermal quench, the electric field becomes sufficiently strong to create a new seed population and drive an avalanche. Figure~\ref{fig:DREAM_TQCQ} shows that a final RE current of $1.15\,\rm MA$ (dash-dotted curve) is predicted in this case (still lower than the predicted $5$-$6\,\rm MA$ with no coil \cite{Tinguely_2021}). The lower logarithmic panel shows that the transport keeps the RE population (thick red dash-dotted line) under control until flux surfaces re-heal in the core around $ 1.6\,\rm ms$ (indicated by dotted line). In the re-healed region the avalanche is effective in multiplying the runaway population to $\sim \rm MA$ level in less than $2\,\rm ms$. Transport data is available from the NIMROD simulation to $3.1\,\rm ms$ (in DREAM simulation time, indicated with vertical dashed line in the upper panel), and they are extrapolated forward such that they do not change significantly in the rest of the simulation, after the flux-surfaces have re-healed. In the log plot of the current, we can see that the total current begins to grow robustly at a time equivalent to $1.8\,\rm ms$ in the NIMROD simulation, corresponding approximately to the time of flux surface reformation evident in the rightmost upper panel of Figure~\ref{fig:SPARC_reheal}. We also note that these results are very conservative with respect to the final runaway current; indeed if we do not employ the mentioned post-processing of the transport coefficients (thin dash-dotted line), the final RE current remains negligible. An important difference between these two treatments is that the post processing removes transport from regions where flux surfaces are intact, but not toroidally symmetric, and as such, particle confinement is not guaranteed on them. We may consider these two results lower and upper as bounds on the RE current.       

Considering the more conservative post-processed transport results, the reformation of flux surfaces is critical for the evolution of the runaway current, and as we have shown, the evolution of the $q$ profile affects when this reformation starts. Even though the $q$-profile evolution in NIMROD cannot account for the RE current component, which it does not calculate, the RE current begins to affect the $q$ profile only when it has reached macroscopic values. Before that, the $q$ profile evolution and corresponding transport obtained from NIMROD can be considered reliable. It is possible to follow the $q$ profile evolution in DREAM, while the simulation presented here does not allow for an ongoing MHD-driven relaxation of the current density profile, and as such it is not expected to be reliable. Without showing a corresponding figure we note that in the DREAM simulation $q(r=0)$ remains below $2$ in the entire simulation. It is clear though that when the RE beam becomes significant, being highly concentrated in the core where the flux surfaces are reformed, the $q$ profile starts to drop rapidly. Such reduction of $q$ as a RE beam begins to form in the core could have a self-mitigating effect as the profile once again becomes resonant with the coil, but further modeling is needed to explore this possibility. A complete model of the $q$ profile evolution would require self-consistent MHD and RE evolution, as well as the inclusion of a resistive wall. We now turn our discussion to the effects of the ideal wall in NIMROD.

\begin{figure}[ht]
    \includegraphics[width=0.48\textwidth]{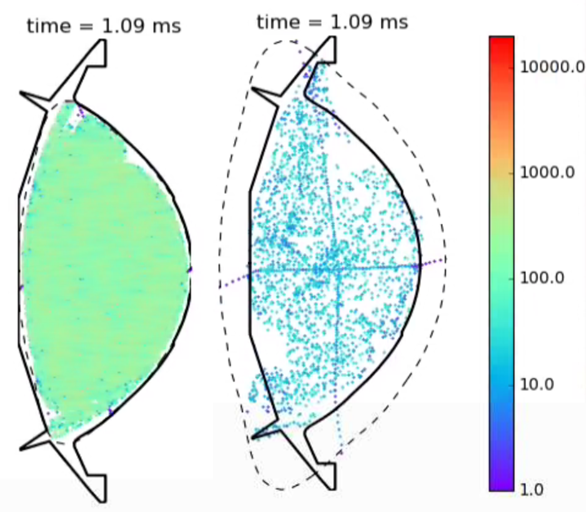}
    \caption{A comparison of field line Poicare plots when the $n=1$ mode reaches maximum amplitude for two TQ-only simulations with simulation boundary (dashed) placed (left) just inside the limiter location (solid) and (right) well outside the limiter (and far from the last closed flux surface). The field lines are colored according to the number of toroidal transits before striking the boundary. The same launch points are used in each case, such that the greater density of points on the left is also a result of longer field lines.
    }
    \label{fig:SPARC_wall}
    \end{figure}

In every simulation that includes the REMC, the NIMROD model requires placement of the ideal wall very near to the last closed flux surface, because the coil itself cannot be accommodated within the simulation domain. Intuitively, the most relevant physical effects of the artificially close conducting wall should be the reduction of the total inductance inside the domain and the stabilization of MHD modes. We can gain some quantitative insight into these effects by comparing the TQ-only simulation that has the same wall shape as the REMC simulations, with a TQ-only simulation having a more distant wall, which is also plotted in Figure \ref{fig:SPARC_n1all}. As expected from the larger inductance, the TQ simulation with a farther wall has a slower CQ. Most importantly, without the stabilizing effect of the close wall, the RE losses begin earlier and reach a higher loss rate, which persists for a longer time-interval. The estimated losses in this case reach $-475$ e-folds--which, while less dramatic than the estimate for the TQ + coil simulation, should still amount to the total loss of the RE seed. A comparison of the wall geometries and of the field lines when maximum stochasticity is achieved is plotted in Figure \ref{fig:SPARC_wall}. Although both simulations show no evidence of good flux surfaces anywhere in the volume when the $n=1$ mode reaches its peak amplitude, when the wall is placed farther out, significantly shorter field lines are observed, leading to the more rapid RE losses. The flux surfaces also take longer to re-heal with the father wall, as evidenced by the longer interval before RE confinement returns.   

\subsection{Variation of the CQ rate and maximum REMC current}

In simulations that include the REMC the CQ rate can not be varied by moving the wall, but the resistivity can be varied to understand the effects of the CQ rate on the coil behavior. The baseline $n=1$ coil simulation was designed to achieve the anticipated maximum CQ rate expected in SPARC, which is $3\,\rm ms$. In both artificial and realistic TQ simulations, the final CQ temperature is determined by the balance of Ne radiation and Ohmic heating and can be varied by changing the Ne quantity. With the Ne density reduced by a factor of four, the core temperature during the TQ increases from $8\,\rm eV$ to $17\,\rm eV$, and the current decay slows accordingly. A slower current decay implies a slower avalanche growth rate but also a slower rise time for the REMC current. A comparison of the $n=1$ coil-only simulations in Figure \ref{fig:SPARC_n1all} reveals that a slower CQ produces overall more optimistic results. Although the RE losses occur later in the absolute time index when the CQ resistivity is decreased, they occur earlier with respect to the avalanche growth-- that is, the number of e-folds predicted prior to the onset of RE losses is smaller. Furthermore, a higher loss rate is found at lower resistivity and the loss fraction is $100\%$. 

    \begin{figure}[ht]
    \includegraphics[width=0.48\textwidth]{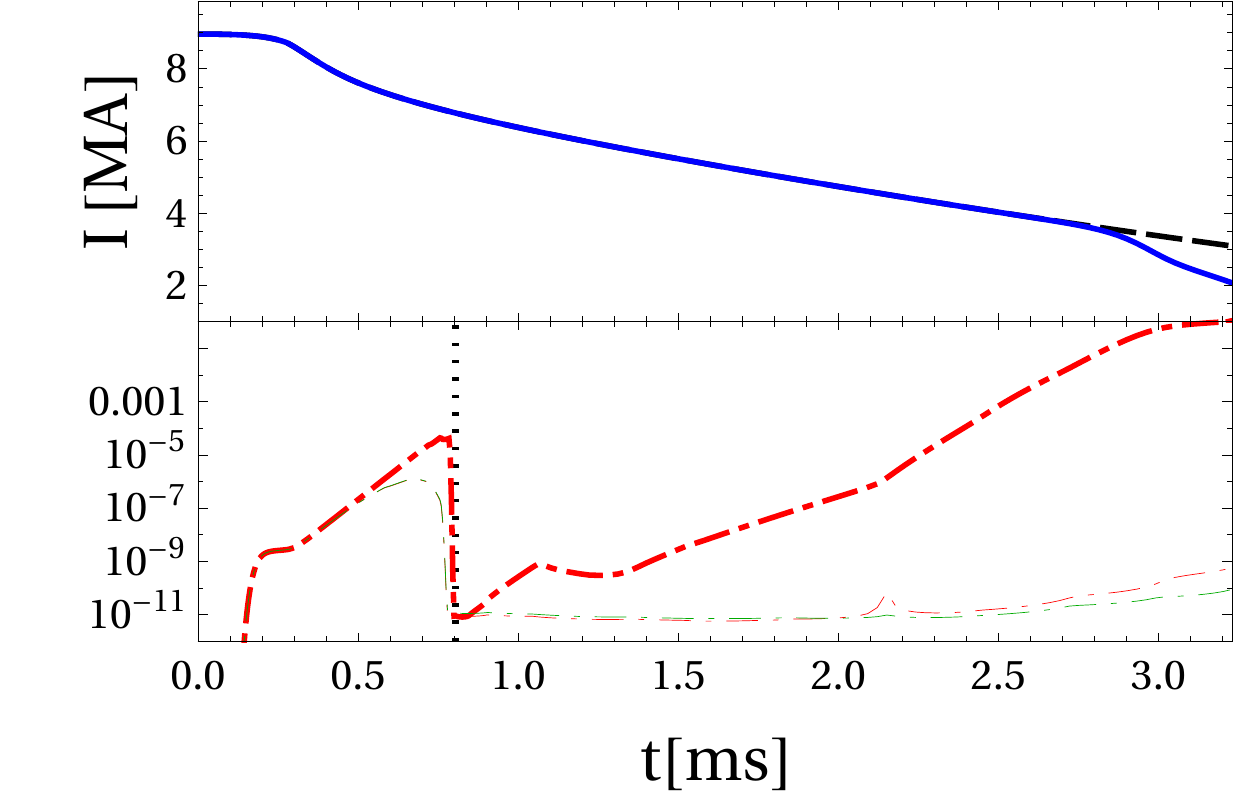}
    \caption{Total current (black dashed), Ohmic current (blue solid), and RE current (thick red dash-dotted), calculated by DREAM simulation with the SPARC REMC current limited to $250\,\rm kA$. Mind the linear (logarithmic) scales on the upper (lower) panel.  All flux surfaces break up for a short while around the time indicated by the dotted vertical line, followed by a rapid reformation of flux surfaces in the core, and corresponding revival of the avalanche growth.  We also show the runaway currents calculated for transport coefficients without post-processing (thin dash-dotted lines), for both the $250\,\rm kA$ (red) and the full coil current (green) cases. 
    }
    \label{fig:DREAM_CQ250kA}
    \end{figure}

The REMC current rise time is governed by the coil inductance, while the low coil resistance produces a near linear relationship between the plasma current and coil current. The addition of some resistance in the coil circuit has been considered to limit the maximum coil current, and therefore the maximum sideways forces produced by the vertical legs of the coil, without significantly affecting the early time coil response. A simulation is performed in which the coil current is clamped at a maximum value of $250\,\rm kA$, although of course the real current wave-form with a resistor would include some rollover prior to reaching $250\,\rm kA$, which is not accounted for here. Because full suppression of the RE current for the coil-only case was previously reported in \cite{Tinguely_2021}, it is desirable to test to limits of that result, although a current-clamped case which includes the TQ-MHD could also be modeled in the future. Note that in the $n=1$ coil-only simulation, the REMC current does not reach $250\,\rm kA$ until $1.33\,\rm ms$, after the period of rapid RE losses. Therefore, the current-clamped simulation is merely a continuation of the coil-only simulation beginning from that time with the coil current held constant thereafter, and the main point of comparison with the original simulation is the rate of flux surface rehealing. We repeat the analysis of calculating the transport coefficients with ASCOT5 and the RE current evolution with DREAM for this case, and find that when the coil current is limited to $250\,\rm kA$ and the small transport levels in the closed flux regions are removed, a late time RE current begins to grow (see Fig.~\ref{fig:DREAM_CQ250kA}).  The final RE current plateau is not reached due to numerical difficulties at the end of the simulation, but the RE current is expected to reach values somewhat above $2\,\rm MA$. Similarly to the results shown in Fig.~\ref{fig:DREAM_TQCQ}, when the transport coefficients are employed without post processing (thin red dash-dotted line), the results are much more favorable, showing the development of a vanishingly small RE current. In fact, these results are comparable to the non-post-processed result with the full REMC current (green dash-dotted line).

    \section{DIII-D modeling}\label{sec:d3d} 

An optimization study for helical coil configurations in DIII-D was reported in Ref. \cite{Weisberg2021}, where the coil configurations considered made a single toroidal turn along the inboard side of the vessel with varying poloidal pitch and had a purely poloidal return path either along the inboard side of the vessel or around the outboard side. Modeling results predicted an induced current in the coil of up to $12\%$ of the pre-disruption plasma current. MHD modeling of the linear plasma response to the coil fields was performed with the MARS-F code for a DIII-D equilibrium limited on the inboard wall, typical of a plasma midway through the CQ. The reconstructed inner-wall-limited (IWL) equilibrium had $q_{95}=4.8$, but a lower-$q$ scenario with the plasma current doubled was also modeled. Using the REORBIT module in MARS-F to calculate RE drift orbits, RE loss fractions of $35$ and $55\%$ (of an initially uniform-in-$\psi$ seed population) were obtained for mid-CQ coil currents of $50\,\rm kA$ and $100\,\rm kA$, respectively, for the higher-$q$ scenario. Larger loss fractions were obtained when $q$ was decreased. 

\begin{figure}[ht]
    \includegraphics[width=0.48\textwidth]{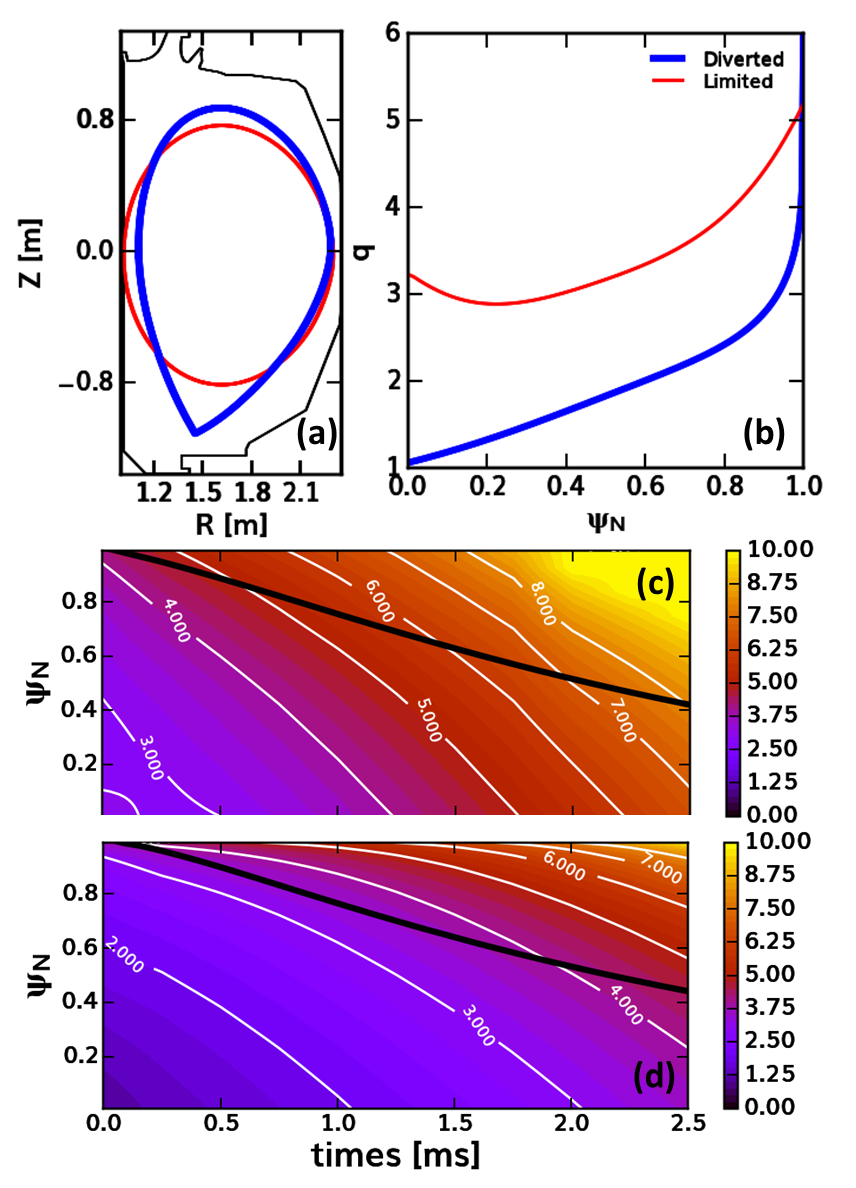}
    \caption{(a) Equilibrium boundary shapes for the DIII-D limited and diverted equilibria. (b) Initial profiles of safety factor, $q$. (c-d) Time evolution of $q$-profile for the (c) limited and (d) diverted plasmas (contours), and normalized plasma current decay (black), where $I_{p,t=0} = 0.81 \,\rm MA$ ($1.46 \,\rm MA$) for the limited (diverted) case.    
    }
    \label{fig:D3Deq}
    \end{figure}

\begin{figure}[ht]
    \includegraphics[width=0.48\textwidth]{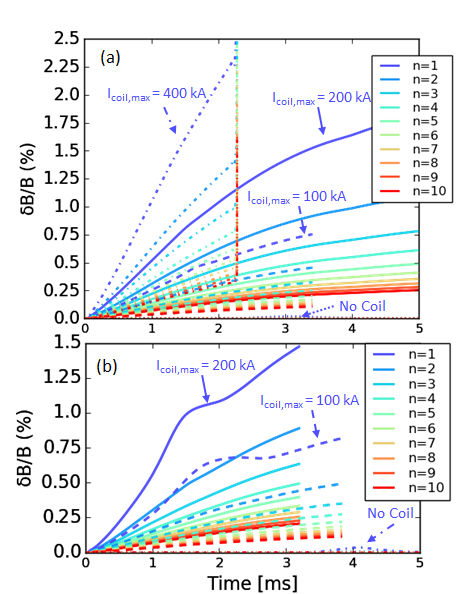}
    \caption{Growth of mode magnetic energy (in units of $\delta B/B$) for (a) IWL and (b) LSN DIII-D simulations, for three values of maximum coil current: $100\,\rm kA$ (dashed), $200\,\rm kA$ (solid), and $400\,\rm kA$ (dash-dot). The $400\,\rm kA$ simulation in (a) ends in numerical instability at $2.2\,\rm ms$.      }
    \label{fig:D3DSpect}
    \end{figure}

While the MARS-F modeling calculated the time-independent linear response during the mid-CQ, here we begin with an IWL early-CQ equilibrium (Figure \ref{fig:D3Deq}) and use NIMROD to calculate the time-dependent nonlinear response as the plasma current decays and the coil current proportionally increases. We use the 3D coil configuration designated as MK2 in Ref.~\cite{Weisberg2021}. As with the SPARC simulations, the coil current increases linearly with the decrease of the plasma current. The DIII-D coil configuration directly perturbs all toroidal mode numbers (both odd and even) with $n=1$ being the largest, as seen in Figure \ref{fig:D3DSpect}. 

We choose maximum coil currents of $100\,\rm kA$ and $200\,\rm kA$ to correspond to the mid-CQ currents of $50\,\rm kA$ and $100\,\rm kA$ considered in the MARS-F simulations. An additional simulation with higher current and a reference simulation with no coil current are also compared. We also consider a lower-single-null (LSN) diverted flat-top equilibrium, in order to make some connection with the SPARC results for a diverted equilibrium. Again RE drift orbits are calculated in each case. 

\subsection{Inner-wall-limited simulations}
\begin{figure}[ht]
    \includegraphics[width=0.48\textwidth]{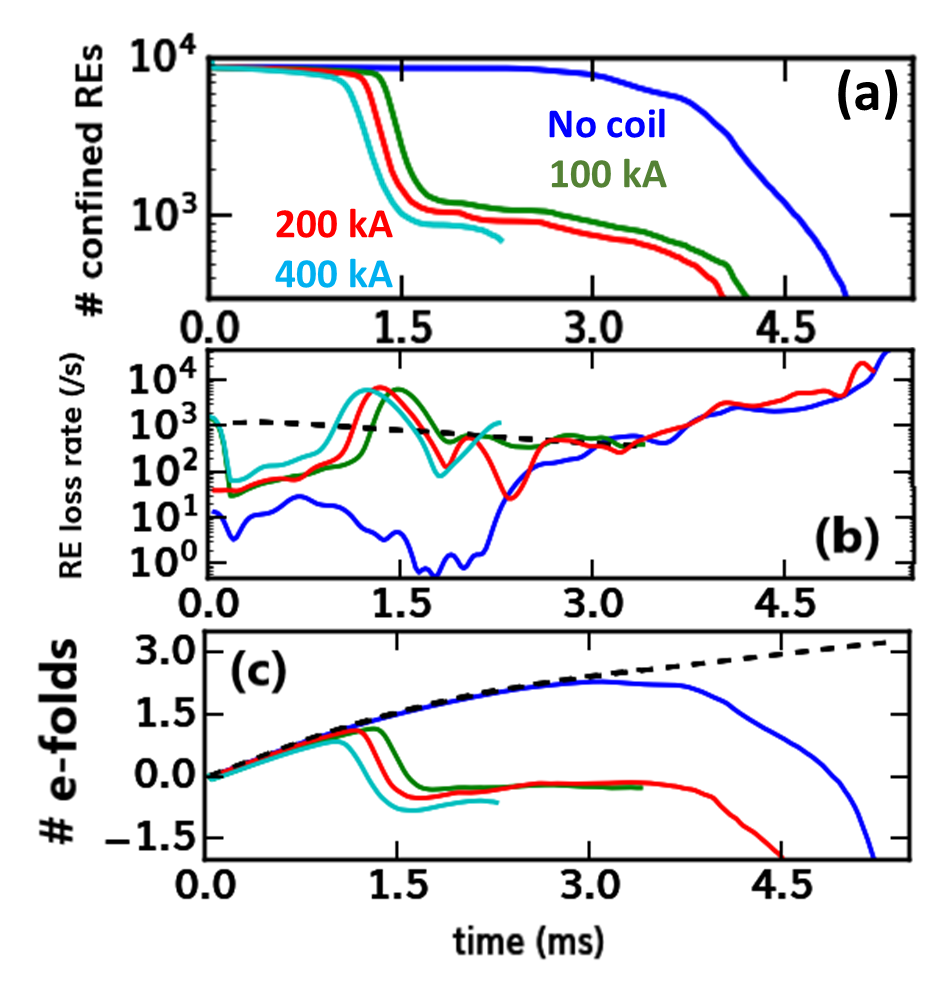}
    \caption{RE losses are insensitive to 3D coil current in the IWL plasma shape. (a) Total number of confined REs vs. time for the limited DIII-D simulations. Even with no coil, the total current is eventually too small to confine any REs. (b) RE loss rates (solid) compared with avalanche growth rate (dashed). (c) Estimated number of avalanche e-folds from integral of growth rate minus loss rate (solid) compared with integrated growth rate (dashed).
    }
    \label{fig:D3DLim}
    \end{figure}

The early-CQ IWL DIII-D equilibrium used as an initial condition has $q_{95}=4.8$ and is shown in Figure \ref{fig:D3Deq} along with the diverted equilibrium. The edge safety factor increases monotonically as the simulation progresses and the plasma current decays. Experimentally, this monotonic increase is not required \cite{Shiraki2016,Paz-Soldan2021} as the plasma can also shrink as it moves into the wall, and a reduction of minor radius is used as an experimental knob to reduce the edge q according to $q_a \propto aB_T /I_P$ \cite{Paz-Soldan2021} in order to trigger kink instabilities, but the ideal-wall boundary condition used in the NIMROD model precludes this effect. Each simulation begins with $10530$ RE test-particles, which are distributed uniformly in normalized poloidal flux (with random poloidal angles) and uniformly in energy ranging from $1$-$50\,\rm MeV$, with pitch ($p_{\perp}/p$) ranging from $0$-$0.5$. In every case, close to $2000$ orbits (initiated near the equilibrium boundary) are lost immediately. In Figure \ref{fig:D3DLim} a comparison of confined REs vs.~time and the corresponding RE loss rate for the IWL case is shown for the cases of no coil perturbation, as well as $100$, $200$, and $400\,\rm kA$ maximum coil current. The most notable feature of these IWL simulations is how insensitive the RE confinement is to the coil amplitude. A rapid loss of test particles is observed beginning just before $1.5\,\rm ms$ in every case except the no-coil simulation, but the primary effect of increasing the coil amplitude is to trigger this loss event just slightly earlier in time. A slight increase in the loss rate prior to the fast loss event is also observed as the coil current increases from $200\,\rm kA$ to $400\,\rm kA$. In every case, the rapid losses begin when the estimated number of avalanche e-folds is $<1.5$ and produce sufficient losses that a final population comparable to or below the seed population is estimated.  The fraction of test particles remaining confined after the rapid loss phase falls between $8$-$12\%$ for the three coil currents.

\begin{figure}[ht]
    \includegraphics[width=0.48\textwidth]{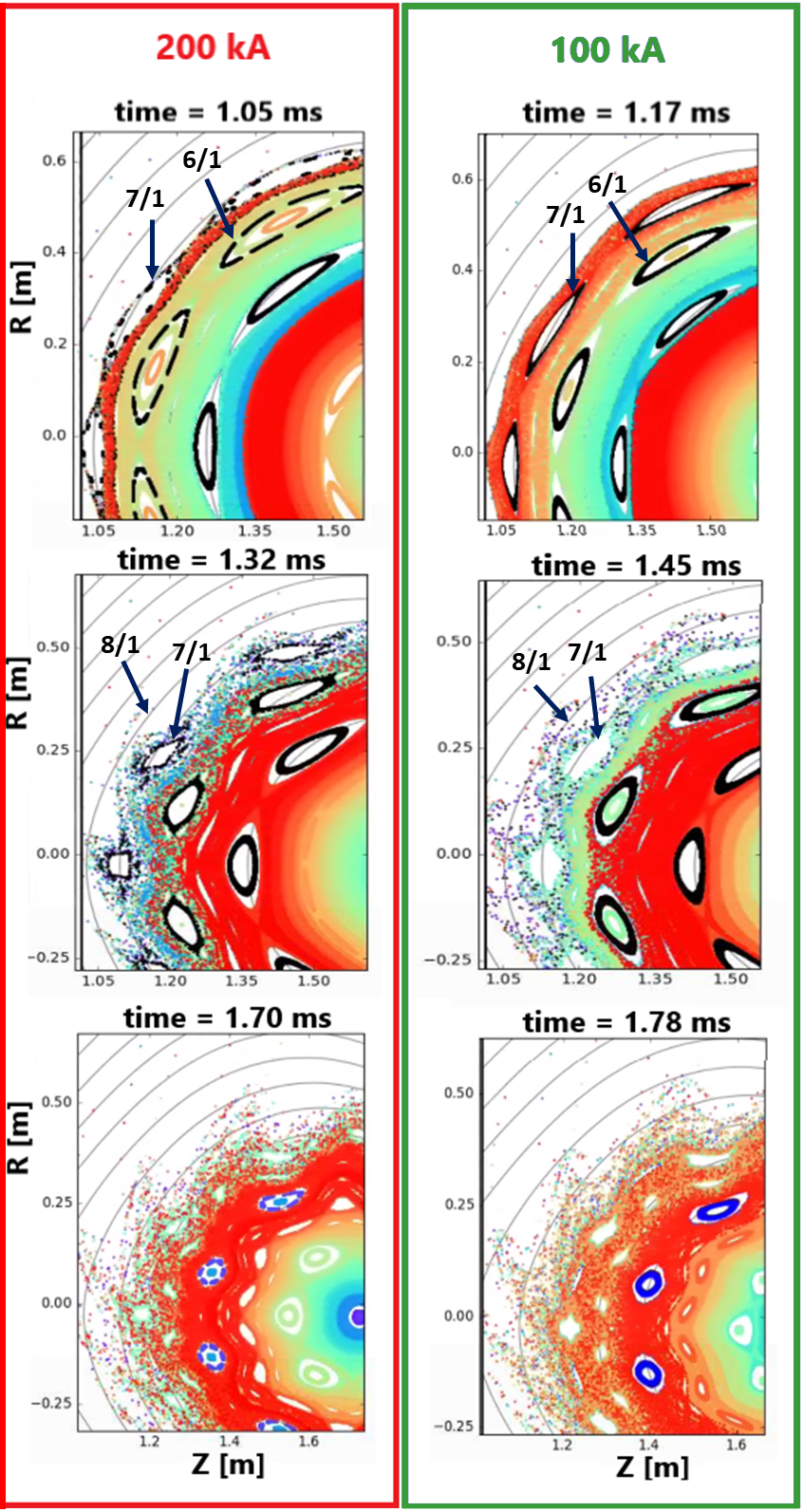}
    \caption{The onset of island overlap and stochasticity is shown in Poincare plots for three times in the $100\,\rm kA$ and $200\,\rm kA$ DIII-D IWL simulations, corresponding to time just before the fast RE losses, at the time of maximum loss rate, and just after the rapid losses have ended.
    }
    \label{fig:D3D_IWL_PSS}
    \end{figure}

    The rapid rise in the edge safety factor compared to the growth rate of the islands from the imposed perturbations is the primary cause of the insensitivity of the RE confinement results to the coil current. As the perturbation amplitude grows, separated islands appear and grow in each case, and as expected these islands are larger when the perturbation amplitude is larger. But the first occurrence of island overlap leading to the onset of stochasticity consistently occurs when the edge safety factor increases above $q=8$, at which time the perturbation amplitude is already sufficient in every case to produce island overlap between the $8/1$ and $7/1$ islands. This triggers an inward cascade of island overlap and a loss of confinement over a significant fraction of the plasma volume (Figure \ref{fig:D3D_IWL_PSS}). Although the growth of the $n=1$ mode above the applied perturbation amplitude is barely visible in Figure \ref{fig:D3DSpect}(a), nonlinear growth and saturation of an $n=1$ mode does occur at the time of the rapid RE losses.   
    
    \subsection{Lower-single-null diverted simulations}
    
      \begin{figure}[ht]
    \includegraphics[width=0.48\textwidth]{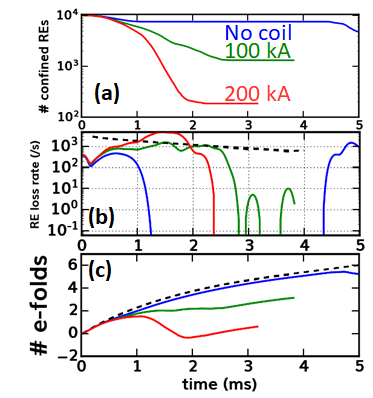}
    \caption{RE loss rate increases with coil current for the LSN plasma shape. (a) Total number of confined REs vs. time for the diverted DIII-D simulations. (b) RE loss rates (solid) compared with avalanche growth rate (dashed). (c) Estimated number of avalanche e-folds from integral of growth rate minus loss rate (solid) compared with integrated growth rate (dashed).
    }
    \label{fig:D3D_REDiv}
    \end{figure}
    
    While the IWL DIII-D simulations were performed to make direct connection to the linear plasma response modeling presented in \cite{Weisberg2021}, in order to make a better connection to the SPARC modeling in a diverted configuration, we also perform simulations of the DIII-D coil operation beginning with a lower single null DIII-D equilibrium having $q$ just above unity on axis, $q_{95}=3.1$, and an initial plasma current of $1.5\,\rm MA$. This LSN equilibrium was also one of a set of six equilibria studied to understand the variability in runaway electron confinement during the TQ in DIII-D \cite{Izzo2012}, with this shot (137611) retaining the most test-particles in the modeling (and having the largest experimental RE plateau) of the set. 
    
    These simulations not only begin with lower safety factor values across the entire radius, but also have more slowly rising $q$-values (Figure \ref{fig:D3Deq}d) relative to the plasma current decay (and coil-current rise), changing the relative importance of the local growth in the resonant field amplitude and the movement of the resonant surfaces. Here we see a marked increase in the RE loss rate as we increase the maximum coil current from $100\,\rm kA$ to $200\,\rm kA$, as well as a significant reduction in the remaining confined fraction when the losses cease (Figure \ref{fig:D3D_REDiv}). The nonlinear growth and saturation of an $n=1$ mode above the applied vacuum fields between $1$ and $2\,\rm ms$ is  more evident from the mode energy spectrum (Figure \ref{fig:D3DSpect})  in these diverted cases than in the limited cases. With no applied external fields an unstable $n=1$ mode only grows late in time, saturating at $\sim 4\,\rm ms$ at a small amplitude of $\delta B/B < 10^{-3}$ (a similarly small mode appears at $\sim 3\,\rm ms$ in the IWL simulation with no coil, although these modes are difficult to see on the mode spectrum plots).   
    
    \begin{figure}[ht]
    \includegraphics[width=0.48\textwidth]{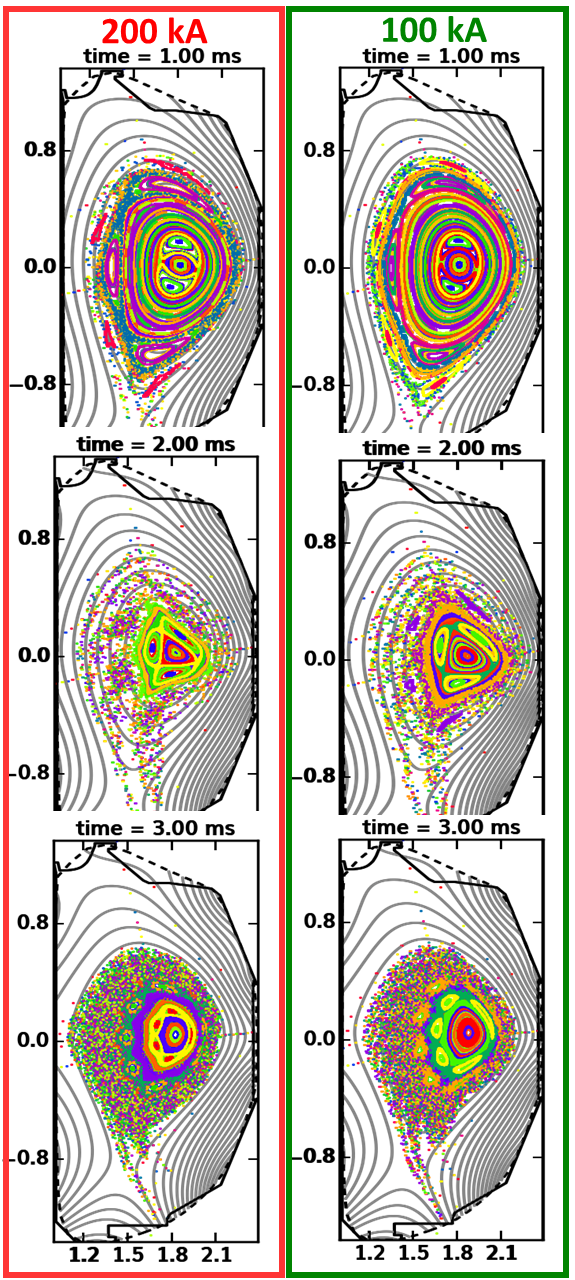}
    \caption{DIII-D LSN simulations with $100kA$ and $200kA$ maximum coil current show destruction of flux surfaces propagating inward as the coil current increases. The safety factor profile evolution is the same in each case so that equivalent island chains at each time can be seen, but with larger islands for the higher current case and a correspondingly smaller region of good flux surfaces. The dashed line is the simulation boundary, which approximately corresponds to the limiter shape (solid). 
    }
    \label{fig:D3DdivPSS}
    \end{figure}
    
    Even with no REMC, about $25\%$ of the REs are lost in the first $1\,\rm ms$ in this case, which includes both an immediate loss of initially unconfined orbits, consisting mainly of highest energy test-particles that strike the outer wall, and a more gradual loss of orbits near the boundary (of all energies) hitting the outer-divertor strike point as the equilibrium responds to the rapid thermal quench. For either level of coil current, the losses in this early phase are faster than with no coil, with losses in the $200\,\rm kA$ simulation becoming increasingly more rapid after $1\,\rm ms$. For a significant interval between $1$ and $2\,\rm ms$, the $200\,\rm kA$ simulation has an RE loss rate exceeding the avalanche growth rate, producing a significant drop in the estimated RE e-folds. Still, by $2\,\rm ms$ the applied fields have reached full penetration with $2\,\rm \%$ of the seed population still confined, although all REs with energy $>30\,\rm MeV$ have been lost. After this time, the particular island chains apparent in the confined region continue to evolve with $q$ (Figure \ref{fig:D3DdivPSS}), but the overall size of the confined region does not shrink further. In contrast, the $100\,\rm kA$ simulation has a RE loss rate that barely equals the avalanche growth rate beginning at $1.5\,\rm ms$ and maintaining that level until $2.5\,\rm ms$, as the stochastic region more slowly penetrates in toward the core. After the losses have ended, $13\%$ of the initial test-particles remain confined, including REs with energies of up to $42\,\rm MeV$. In fact, the residual population in the $200\,\rm kA$ simulation has an average RE energy of $9\,\rm MeV$, while the larger residual population in the $100\,\rm kA$ simulation has an average energy of $16\,\rm MeV$. For the modest avalanche growth in DIII-D, where the size of the initial seed significantly affects the final RE current, even the global estimate of roughly $2$ e-fold reduction in the seed population at $100\,\rm kA$ would be expected to produce a measurable difference in the final RE current.

    \section{Discussion}\label{sec:discussion} 

Massive material injection strategies, such as shattered pellet injection (SPI) or massive gas injection (MGI), have not been satisfactorily demonstrated as a means to prevent RE beam formation in high-current tokamaks. Indeed, as the recent review of Breizman concludes: "[w]ith ITER construction in progress, reliable means of RE mitigation are yet to be developed" \cite{Breizman_2019}. While pre-TQ injection of SPI or MGI remain essential parts of an overall disruption mitigation strategy, given their doubtful prospects for preventing RE beam formation, recent efforts have focused on strategies to mitigate already formed RE beams, such as by secondary SPI injection combined with naturally occurring kink-instabilities in the RE plateau \cite{Paz-Soldan2021,reux2021demonstration}. A passively driven 3D coil could provide an additional tool for RE prevention, to be used in conjunction with SPI or MGI for disruption mitigation. The primary purpose of such a coil would be to deconfine seed REs faster than they can give birth to secondaries, avoiding avalanche multiplication. But in light of recent experiments demonstrating the importance of large magnetic fluctuations for benign RE plateau termination \cite{Paz-Soldan2021,reux2021demonstration}, the coil could be a useful tool even in the event of RE beam formation.

Passively driven coils to mitigate REs have been designed for SPARC and DIII-D to gain knowledge of the operation of such coils for future reactor-scale tokamaks. The SPARC device will also employ MGI as its primary disruption mitigation strategy \cite{sweeney_2020}, while present DIII-D disruption mitigation experiments are focused on (but not limited to) SPI injection into all phases of the disruption (pre-TQ, CQ, or RE plateau) \cite{Shiraki2016,shiraki2018dissipation,Paz-Soldan2021}. The nonlinear-MHD  passive-coil modeling for each device favors the successful operation of each coil for its intended purpose. 

The $n=1$ coil design for SPARC comes with engineering challenges related to sideways forces. But with all three coils having predominantly $m=1$ poloidal symmetry, the $n=2$ and $n=3$ REMC designs for SPARC are not found to have strong enough resonant components to produce RE loss rates that exceeded the avalanche growth rate. Modeling of the $n=1$ REMC in SPARC for a single scenario  with a maximum coil current of $590\,\rm kA$ (using a combination of NIMROD, ASCOT5, and DREAM) was previously shown to predict full suppression of the RE current \cite{Tinguely_2021}. Here, we model the scenario in which the coil current is clamped at 250kA. Following the same procedure, the clamped current case predicts an RE current level that is higher, but still extremely negligible. When the ASCOT5 results were post-processed with much more conservative assumptions, a RE beam was able to develop.

In additional modeling we have explored the effects of some simplifying approximations in the published reference scenario. In particular, while the neglect of the TQ-induced MHD in the work of Tinguley, et al \cite{Tinguely_2021} was intended to isolate the effect of the coil, we find here that the coil and TQ MHD modes interact at the time of the $I_p$-spike to produce significantly faster RE losses than what is seen when either effect is isolated. We also find that the close, ideal wall in the simulations has a strong stabilizing effect on TQ-triggered MHD modes, and by inference, should have a similar effect on those triggered by the coil, although the location of coil precludes moving the wall outward in those simulations. The simulations should be extended to include a resistive rather than an ideal wall at the limiter location. The resistive wall model that has been employed in the NIMROD code for disruption simulations \cite{artola20213d} grids an external vacuum region that is coupled to the plasma region at the resistive-wall boundary, which would also be \textit{prima facie} incompatible with the presence of the 3D coil. A less well-exercised resistive-wall model using a Green's function approach at the first-wall boundary (and no external region) \cite{wang2018simulation} could be employed in the future to improve the fidelity of the present work.

One feature of the TQ + coil modeling was found to be less favorable for RE suppression than the results obtained with the coil-only modeling--namely the rate at which large regions of good flux surfaces reappear. This difference is inherently connected to the much more rapid losses at the time of the $I_p$-spike, because that same large MHD event is also responsible for rearranging the current density profile such that $q$ increases above $2$ on-axis within a short time. After the post-processing of the ASCOT5 results, DREAM predicts the formation of a $1.15\,\rm MA$ RE beam once the flux surfaces reappear, although the RE current again remains negligible when the post-processing is not applied. We have already noted at least one complication in modeling this case with DREAM: due to the practical need for a monotonic current evolution in the DREAM simulation, and the lack of natural MHD-driven current relaxation, some important loss processes during the TQ may not be accounted for. Clearly, the evolution of the safety-factor profile is key to the performance of the coil at later times, so we reiterate that two very important effects that will govern the evolution of the $q$-profile during the CQ are neglected in NIMROD: the resistive wall, and the effects of the RE current itself. The evolution of $q$-on-axis in DREAM indicates that once the REs begin to avalanche in a re-formed core of good flux surfaces, a modest level of current compared to the Ohmic current ($\sim 0.5 \,\rm MA$) would bring $q$ in that region back down toward resonance with the coil, so that subsequent MHD events could limit the growth of the current. This hypothesis could be explored with MHD modeling that includes a fluid RE model, such as that used in JOREK \cite{bandaru2019simulating,bandaru2021magnetohydrodynamic} and M3D-C1 \cite{zhao2020simulation}, with a similar implementation for NIMROD in progress \cite{sainterme2020development}. The possibility of exciting later-time MHD events is a further argument against limiting the maximum current in the REMC by additional resistance unless strongly necessitated by engineering requirements.

Numerous experiments with post-disruption RE current plateaus (the largest carrying almost 1MA of current) have been performed on DIII-D \cite{Eidietis2012,Hollmann2013,shiraki2018dissipation,Paz-Soldan2021} without causing significant damage to the device, making it a valuable test-bed for RE mitigation concepts.  The DIII-D modeling of the IWL plasmas showed a larger loss fraction ($\sim 90\%$) than the linear plasma response modeling from Ref. \cite{Weisberg2021} for equivalent coil current, suggesting that nonlinear excitation of modes enhances losses in this scenario. Beginning with a LSN equilibrium, the loss fraction at $100\,\rm kA$ was comparable to the IWL simulations, but increased to $98\%$ at $200 \,\rm kA$. In part, this may be attributed to the lower safety factor, as losses were also enhanced in Ref. \cite{Weisberg2021} by reducing $q$ in an IWL configuration. However, a diverted plasma is also expected to more easily form a stochastic layer in the presence of externally applied fields, due to larger magnetic shear at the boundary \cite{Evans2002}. 

Even for the most comparable diverted plasma simulations, a number a differences exist between the SPARC and DIII-D cases that could contribute to the greater RE loss fractions seen in SPARC, even for similar values of $I_{REMC}/I_p$. The coil designs differ considerably--one outboard and one inboard--with the DIII-D design having been optimized for the case of an IWL plasma typical of the CQ phase and not for the diverted plasma. As noted in the introduction, the shorter Alfv\'en time in SPARC leads to faster growth of MHD instabilities. Additional differences between the particular equilibria chosen could also contribute and further study would be needed to establish more clearly the relative importance of these factors.

With moderate avalanche gain factors in DIII-D of $50$-$150$, deconfinement of $\sim 90$-$98\%$ of the seed population would almost certainly reduce the final current carried by a post-CQ RE beam, if not eliminate it all together. These modeling results then, in conjunction with the findings of Weisberg \textit{et al.} \cite{Weisberg2021}, serve to bolster the prospect that installation of such a coil on DIII-D will be valuable as an experimental tool to investigate the REMC operation, even if higher loss fractions, such as those predicted for SPARC, are needed for RE suppression in a higher current tokamak.

    \bigskip

\noindent \textbf{Acknowledgements.} The authors are grateful to M. Hoppe and T. F\"{u}l\"{o}p for fruitful discussions. This work is supported in part by Commonwealth Fusion Systems. This work was partly supported by the Swedish Research Council (Dnr. 2018-03911). This material is also based on work supported by the Department of Energy under Awards Number DE-FG02-95ER54309 and DE-FC02-04ER54698. This research used resources of the National Energy Research Scientific Computing Center (NERSC), a U.S. Department of Energy Office of Science User Facility operated under Contract No. DE-AC02-05CH11231. Part of the data analysis was performed using the OMFIT integrated modeling framework \cite{Meneghini2015}. 

\noindent \textbf{Disclaimer.} Part of this report was prepared as an account of work sponsored by an agency of the United States Government. Neither the United States Government nor any agency thereof, nor any of their employees, makes any warranty, express or implied, or assumes any legal liability or responsibility for the accuracy, completeness, or usefulness of any information, apparatus, product, or process disclosed, or represents that its use would not infringe privately owned rights. Reference herein to any specific commercial product, process, or service by trade name, trademark, manufacturer, or otherwise does not necessarily constitute or imply its endorsement, recommendation, or favoring by the United States Government or any agency thereof. The views and opinions of authors expressed herein do not necessarily state or reflect those of the United States Government or any agency thereof.
    
    \bigskip
    
    \bibliographystyle{unsrt}
    
    \bibliography{bib}

\end{document}